\documentclass[twocolumn,showpacs,preprintnumbers,amsmath,amssymb]{revtex4}
\usepackage{amsmath}
\usepackage{bm}
\usepackage{epsfig}
\usepackage[colorlinks=true,linkcolor=blue]{hyperref}

\newcommand{\be}{\begin{equation}}
\newcommand{\ee}{\end{equation}}

\newcommand{\bea}{\begin{eqnarray}}
\newcommand{\eea}{\end{eqnarray}}

\newcommand{\al}{\alpha}
\newcommand{\bt}{\beta}
\newcommand{\gm}{\gamma}
\newcommand{\Gm}{\Gamma}
\newcommand{\dl}{\delta}
\newcommand{\Dl}{\Delta}
\newcommand{\eps}{\epsilon}

\newcommand{\et}{\eta}

\newcommand{\lm}{\lambda}

\newcommand{\rh}{\rho}

\newcommand{\ta}{\tau}

\newcommand{\om}{\omega}
\newcommand{\Om}{\Omega}

\newcommand{\nn}{\nonumber}

\begin{document}

\title{Excitation of MHD waves in magnetized anisotropic cosmologies}

\author{Apostolos Kuiroukidis$^{1,2}$, Kostas Kleidis$^{1,3}$, Demetrios Papadopoulos$^1$
and Loukas Vlahos$^1$}

\affiliation{$^1$Department of Physics, Aristotle University of
Thessaloniki, 54124 Thessaloniki, Greece}

\affiliation{$^2$Department of Informatics, Technological
Education Institute of Serres, 62124 Serres, Greece}

\affiliation{$^3$Department of Civil Engineering, Technological
Education Institute of Serres, 62124 Serres, Greece}

\date{\today}

\begin{abstract}
The excitation of cosmological perturbations in an anisotropic
cosmological model and in the presence of a homogeneous magnetic
field was studied, using the resistive magnetohydrodynamic (MHD)
equations. We have shown that fast-magnetosonic modes, propagating
normal to the magnetic field grow exponentially and {\em
saturated} at high values, due to the resistivity. We also
demonstrate  that the jeans-like instabilities are enhanced inside
a resistive and the formation of condensations formed within an
anisotropic fluid influence the growing magnetosonic waves.
\end{abstract}


\maketitle

\section{Introduction}

Magnetic fields are known to have a widespread presence in our
Universe, being a common property of the intergalactic medium in
galaxy clusters (Kronberg 1994), while, reports on Faraday
rotation imply significant magnetic fields in condensations at
high redshifts (Kronberg, Perry \& Zukowski 1992). Studies of
large-scale magnetic fields and their potential implications for
the formation and the evolution of the observed structures, have
been the subject of continuous theoretical investigation (see
Thorne 1967, Jacobs 1968, Ruzmaikina \& Ruzmaikin 1971, Wasserman
1978, Zel'dovich, Ruzmaikin \& Sokoloff 1983, Adams, Danielsson \&
Rubinstein 1996, Barrow, Ferreira \& Silk 1997, Tsagas \& Barrow
1997, Jedamzik, Katalinic \& Olinto 1998, Barrow, Maartens \&
Tsagas 2006, for a representative, though incomplete list).
Magnetic fields observed in galaxies and galaxy clusters are in
energy equipartition with the gas and the cosmic rays (Wolfe,
Lanzetta \& Oren 1992). The origin of these fields, which can be
astrophysical, cosmological or both, remains an unresolved issue.

If magnetism has a cosmological origin, as observations of $\mu$G
fields in galaxy clusters and high-redshift protogalaxies seem to
suggest, it could have affected the evolution of the Universe
(Giovannini 2004, Barrow, Maartens \& Tsagas 2006). There are
several scenarios for the generation of primordial magnetic fields
(e.g. see Grasso \& Rubinstein 2001). Most of the early treatments
were Newtonian, with the relativistic studies making a relatively
recent appearance in the literature. A common factor between
almost all the approaches is the use of the MHD approximation,
namely the assumption that the magnetic field is frozen into an
effectively infinitely conductive cosmic medium (i.e. a fluid of
zero resistivity). With a few exceptions (e.g. see Fennelly 1980,
Jedamzik, Katalinic \& Olinto 2000, Vlahos, Tsagas \& Papadopoulos
2005), the role of kinetic viscosity and the possibility of
non-zero resistivity have been ignored. Nevertheless, these
aspects are essential for putting together a comprehensive picture
of the magnetic behavior, particularly as regards the non-linear
regime. The electric fields associated with the resistivity can
become the source for particle acceleration, while the induced
non-linear currents may react back on the magnetic field (Vlahos,
Tsagas \& Papadopoulos 2005).

Many recent studies have used a Newtonian or a Friedmann -
Robertson - Walker (FRW) model to represent the evolving Universe
and super-imposed a large-scale ordered magnetic field. The
magnetic field is assumed to be too weak to destroy the FRW
isotropy and the anisotropy, induced by it, is treated as a
perturbation (Ruzmaikina \& Ruzmaikin 1971, Tsagas \& Barrow 1997,
Durrer, Kahniashvil \& Yates 1998). Current observations give a
strong motivation for the adoption of a FRW model but the
uncertainties on the cosmological {\em Standard Model} are
several. Therefore, the limits of the approximations and the
effects one may lose by neglecting the anisotropy of the
background magnetic field, should be investigated. Within this
context, the formation of small-scale structures and the
excitation of resistive instabilities in Bianchi-Type models has
been explored several years ago (Fennelly 1980). Nevertheless, the
excitation of MHD-waves in curved spacetime and their subsequent
temporal evolution, is far from being exhausted (Papadopoulos,
Vlahos \& Esposito 2001).

In the present article we explore the evolution of a magnetized
resistive plasma in an anisotropic cosmological model. We begin
with a uniform plasma driving the dynamics of the curved spacetime
(the so-called {\em zeroth-order solution}). This dynamical system
is subsequently perturbed by small-scale fluctuations and we study
their interaction with the anisotropic background, searching for
imprints on the temporal evolution of the perturbations'
amplitude. In particular:

In Section II, we present the system of the field equations
appropriate to describe the model under consideration. In Section
III, we solve this system analytically, to derive the zeroth-order
solution. Accordingly, in Section IV, we extract the first-order
perturbed equations. In Section V, we derive the dispersion
relation for the magnetized cosmological perturbations and in
Section VI, we perform a numerical study of their evolution, using
a fifth-order Runge - Kutta - Fehlberg temporal integration
scheme. Finally, in Section VII, a perturbation analysis over
purely gravitational fluctuations reveals an inherent Jeans-like
instability.

Our results suggest that in a resistive plasma, within an interval
of $10^{11} sec$ after the beginning of the interaction process,
fast-magnetosonic modes are excited, growing exponentially in time
and is saturated at high values. In this way, magnetic field
perturbations can be retained at large amplitudes, for
sufficiently long time-intervals $(\Dl t \sim 10^{12} \; sec)$,
resulting in the enhancement of the ambient magnetic field (dynamo
effect). In addition, the resistive plasma enhance the
condensations that can be formed within the anisotropic fluid due
to a gravitational instability, which, in turn, influence the
growth of the magnetosonic waves.

\section{The field equations}

We consider an axisymmetric Bianchi-Type I cosmological model,
driven by an anisotropic and resistive perfect fluid, in the
presence of a time-dependent magnetic field, ${\vec B} = B(t)
{\hat x}$. The corresponding line-element is written in the form
\be ds^2 = - dt^2 + R^2 (t) dx^2 + S^2 (t) [dy^2 + dz^2] \ee The
evolution of a curved spacetime in the presence of matter and an
e/m field, is determined by the gravitational field equations \be
{\cal R}_{\mu \nu} - {1 \over 2} g_{\mu \nu} {\cal R} = + 8 \pi G
{\cal T}_{\mu \nu} \ee (in the system of units where $\hbar = 1 =
c$), together with the energy-momentum conservation law \be {\cal
T}_{ \; \; ; \nu}^{\mu\nu} = 0 \ee and Maxwell's
equations \bea F_{\; \; ;\nu}^{\mu \nu} & = & 4 \pi J^{\mu}, \\
F_{\mu \nu ; \lm} & + & F_{\nu \lm ; \mu} + F_{\lm \mu ; \nu} = 0
\eea In Eqs (2) - (5), Greek indices refer to the four-dimensional
spacetime (in accordance, Latin indices refer to the
three-dimensional spatial section) and the semicolon denotes
covariant derivative. Furthermore, ${\cal R}_{\mu\nu}$ and ${\cal
R}$ are the Ricci tensor and the scalar curvature with respect to
the background metric $g_{\mu \nu}$, while $G$ is Newton's
gravitational constant. Eventually, $F^{\mu \nu}$ is the
antisymmetric tensor of the e/m field and $J^{\mu}$ is the
corresponding current density.

The energy-momentum tensor involved, consists of two parts; namely
\be {\cal T}^{\mu \nu} = {\cal T}_{fluid}^{\mu \nu} + {\cal
T}_{em}^{\mu \nu}\ee The first part, is due to an anisotropic
perfect fluid source, of the form \be {\cal T}^{\mu \nu}_{fluid}=
\rh u^0 u^0 + p_i u^i u^i + p_i g^{ii} \ee where, $\rh(t)$ is the
energy density, $p_i(t)$ are the components of the anisotropic
pressure and the axial symmetry of the metric (1) implies that
$p_2 (t) = p_3 (t)$. Finally, $u^{\mu} = dx^{\mu}/ds$ is the
fluid's four-velocity, satisfying the conditions $u_{\mu} u^{\mu}
= -1$ and $h^{\mu \nu} u_{\mu} = 0$, with $h^{\mu \nu} = g^{\mu
\nu} + u^{\mu} u^{\nu}$ being the projection tensor.

The second part, is due to the ambient e/m field \be {\cal T}^{\mu
\nu }_{em}= {1 \over 4 \pi} ( F^{\mu \al} F^{\nu \bt} g_{\al \bt}
- \frac{1}{4} g^{\mu \nu} F_{\al \bt} F^{\al \bt}) \ee where, for
${\vec E} = {\vec 0}$ and ${\vec B}$
// ${\hat x}$, the non-zero components of the Faraday tensor in
the curved spacetime (1) read (see Appendix A) \be F^{23}= {B^x
\over S^2} = -F^{32} \ee As regards the current density $J^{\mu}$,
it may be determined by the invariant form of Ohm's law \be
J^{\mu} = n_e e u^{\mu} + {1 \over \et} F^{\mu \nu} u_{\nu} \ee
where, $n_e$ is the locally measured charge density and $\et$ is
the (finite) electric resistivity, in units of {\it time}. As a
consequence of the Maxwell equations, we obtain $J_{; \mu}^{\mu} =
0$. Assuming that that the fluid has zero net-charge, i.e.
$n_e=0$, Eq (10) reduces to $J^{\mu} = {1 \over \et} F^{\mu \nu}
u_{\nu}$. A vanishing net charge indicates that the perfect fluid
consists, at least, of two components.

\section{The background solution}

We look for axisymmetric Bianchi-Type I cosmological solutions to
the Einstein-Maxwell equations (Appendices A and B), representing
the background metric of our problem. In this case, Eqs (2) reduce
to \bea &&2 \left(\frac{\dot{R}\dot{S}}{R S}\right) +
\left(\frac{\dot{S}}{S}\right)^{2} = 8 \pi G
\rh(t) + G B^2(t)\nn \\
&&-2\frac{\ddot{S}}{S} - \left(\frac{\dot{S}}{S}\right)^{2}
= 8 \pi G p_1(t) - G B^2(t)\\
&&-\left(\frac{\ddot{R}}{R} + \frac{\ddot{S}}{S}\right)
-\left(\frac{\dot{R}\dot{S}}{R S}\right) = 8 \pi G p_2(t) + G
B^2(t) \nn \eea (the dot denotes time-derivative) and Eqs (4), (5)
yield \be \partial_{t}[S^2 B(t)] = 0 \ee Eq (12) has a clear
physical interpretation: The magnetic flux through a comoving
surface normal to the direction of the magnetic field, {\em is
conserved}.

On the other hand, the continuity equation (3) results in
(Appendix C) \bea &&\partial_t [\rh(t) + {B^2 \over 8 \pi}] +
{\dot{R} \over R} \: [ p_1(t) - {B^2 \over 8 \pi}] + 2 {\dot{S}
\over S} \: [ p_2(t) + {B^2 \over 8 \pi} ] \nn \\
&&+ ({\dot{R} \over R} + 2 {\dot{S} \over S}) \: [\rh (t) + {B^2
\over 8 \pi}] = 0 \eea and the particles' number conservation law,
reads \be {\dot{\rh}} + (\frac{\dot{R}}{R} + 2 \frac{\dot{S}}{S})
\rho = 0 \ee The system of Eqs (11) - (14) admits the exact
solution \bea R(t) = ({t \over t_0}), \; \; \; \; S(t) = ({t \over
t_0})^{1 \over 2} \nn \\ \rh (t) = \rh_0 ({t_0 \over
t})^2 \; \; \; \; B (t) = B_0 ({t_0 \over t}) \\
p_1 (t) = p_{10} ({t_0 \over t})^2 , \; \; \; p_2 (t) = - p_{20}
({t_0 \over t})^2 \nn \eea where, the index "$0$" stands for the
corresponding values at $t = t_0$ and $t_0$ marks the beginning of
the interaction between magnetized plasma and curved spacetime.
Solution (15), represents an anisotropic cosmological model, in
which, the large-scale anisotropy along the $\hat{x}$-axis, is due
to the presence of an ambient magnetic field. The combination of
Eqs (11) and (15) indicates that, initially, the total energy
density is given by \be \rho_0 + {B_0^2 \over 8 \pi} = {5 \over 32
\pi G t_0^2} \ee and the difference between fluid's pressure and
the pressure of the magnetic field along the two anisotropic
spatial directions is equal \bea p_{10} - {B_0^2 \over 8 \pi}
& = & {1 \over 32 \pi G t_0^2} \nn \\
p_{20} - {B_0^2 \over 8 \pi} & = & {1 \over 32 \pi G t_0^2} \eea
Eqs (17) lead us to identify \be p_{10} = p_0 = p_{20} \ee i.e.
initially, when $R(t_0) = S(t_0)$, the two components of the
anisotropic pressure were equal in absolute value, something that
is confirmed also by Eq (13). Furthermore, with the aid of Eqs
(16) and (17), we obtain \be p_{10} + 2 p_{20} = {3 \over 5}
(\rho_0 + 6 {B_0^2 \over 8 \pi}) \ee which, according to Eq (18),
results in the equation of state for the matter-energy content at
$t = t_0$; namely, \be p_0 = {1 \over 5} (\rho_0 + 6 {B_0^2 \over
8 \pi}) \ee For $B_0 = 0$, i.e. as regards the perfect fluid
itself, we obtain that, initially, $p_0 = {1 \over 5} \rh_0$.
Since $p_0 < {1 \over 3} \rh_0$, our model corresponds to a {\em
semi-realistic} cosmological model of Bianchi Type I. These models
are crude, first order approximations to the actual Universe when
we use currently available theories and observations (Jacobs
1969).

\section{The cosmological perturbations}

For any dynamical system, much can be learnt by investigating the
possible modes of small-amplitude oscillations or waves. A plasma
is physically much more complicated than an ideal gas, especially
when there is an externally applied magnetic field. As a result, a
variety of small-scale perturbations may appear. We first assume a
uniform magnetized plasma in curved spacetime as background, which
is perturbed by small scale fluctuations. In this article, the
evolution of the background is described by the solution (15).

Accordingly, we introduce first-order perturbations to the
Einstein-Maxwell equations, by decomposing the physical variables
of the fluid as \bea \rho (t,z) & = & \rho (t) + \dl
\rho(t,z) \\
p_x (t,z) & = & p_1 (t) \nn \\
p_y (t,z) & = & p_2 (t) - \dl p (t,z) \\
p_z (t,z) & = & p_2 (t) + \dl p (t,z) \nn \eea and we insert the
perturbed values (21) and (22) into Eqs (11) - (14), neglecting
all terms higher or equal than the second order. The pressure
perturbation $\dl p (t, z)$ introduces a longitudinal acoustic
mode, propagating along the ${\hat z}$-direction and therefore \be
\dl p (t, z) = C_s^2 \: \dl \rh (t, z) \ee where, $C_s$ is the
speed of sound. The four-velocity of the plasma fluid is perturbed
around its comoving value, $u^{\mu} = (1,0,0,0)$, as \be u^{\mu}
(t,z) = (1+\dl u^{0}(t,z),0,0,\dl u^{z}(t,z)) \ee Then, the
condition $u_{\mu} u^{\mu} = -1$, to the first leading order,
implies \be \dl u^{0}(t,z)=0 \ee and, therefore, $u^3(t, z) = \dl
u^z (t, z)$. Accordingly, $\rho (t,z)u^3(t,z) = \rho (t) \dl
u^z(t,z) + O_2$.

As regards the perturbations of the e/m field, we consider that
they correspond to a transverse e/m wave, propagating along the
${\hat z}$-axis $({\vec k} // {\hat z})$; namely, \bea
\vec{E}(t,z) & = &\dl E^{y}(t,z)\hat{y} \\
\vec{B}(t,z)&=&B(t)\hat{x}+\dl B^{x}(t,z)\hat{x} \eea Therefore,
now, the non-zero components of the Faraday tensor in curved
spacetime are modified as follows \bea && F^{02}= {1 \over S} \:
\dl E^y (t, z) = - F^{20} \nn \\
&& F^{23} = {1 \over S^2} \: [B(t) + \dl B^x(t, z)] = - F^{32}
\eea In what follows, we take into account the so-called {\em
Cowling approximation} (Cowling 1941), admitting that $\dl g_{\mu
\nu} = 0$. Therefore, the evolution of the perturbed quantities is
governed only by the energy-momentum tensor conservation, together
with Maxwell's equations.

To begin with, we perturb the particles' number conservation law:
Accordingly, Eq (C3), yields \be \partial_t (\dl \rho ) + \rho(t)
\partial_z (\dl u^z) + \dl \rho \left(\frac{\dot{R}}{R} +
2\frac{\dot{S}}{S}\right) = 0 \ee We continue with Maxwell's
equations: Then, from Eq (B2), using Eqs (21), (22), (24) and
(25), we obtain \bea -\partial_t (\dl E^{y}) & + & {1 \over S}
\partial_z (\dl B^x) - \dl E^{y} \left ( \frac{\dot{R}}{R} +
\frac{\dot{S}}{S} \right) = \nn
\\ & = & 4 \pi {1 \over \et} [\dl E^{y} + S B (t) \dl u^{z}] \eea
Now, Eq (B3) becomes \be \partial _{t}(S^2 \dl B^{x}) - S
\partial_{z} (\dl E^{y}) = 0 \ee Eventually, the conservation
equation (C2) results in \bea &&\partial_{t} [\rho(t)\dl u^{z} -
{1 \over 4 \pi S} B(t) \dl E^{y}] + {1 \over S^2} \partial_{z}
[\dl p + {1 \over 4 \pi} B(t) \dl B^{x}]+ \nn \\
&&+\left(\frac{\dot{R}}{R}+2\frac{\dot{S}}{S}\right) [\rho(t) \dl
u^{z} - {1 \over 4 \pi S} B(t) \dl E^{y}]=0 \eea while, to the
first leading order, Eq (C1) collapses to an identity. Eqs (29) -
(32) are the linearly independent first order perturbed
Einstein-Maxwell equations in the curved background (1), which we
intend to discuss. Notice that, in the flat spacetime - zero
resistivity limit, they reduce to Eqs (10.53a), (10.9) and
(10.53c) of (Jackson 1975), respectively.

To develop the theory of small-amplitude waves in curved
spacetime, we search for solutions to the linearized equations
(29) - (32), in which all perturbation quantities are proportional
to the exponential \be \exp \: [i(k z - \int^t \om dt)] \ee
following the so-called {\em adiabatic approximation} (Zel'dovich
1979, Birrell \& Davies 1982, Padmanabhan 1993). In this context,
the (slowly varying) time-dependent frequency of the wave is
defined by the {\em eikonal} \be \Om = \int^t \om dt \ee through
the relation \be \om = {d \Om \over dt} \ee Notice that, in Eq
(33), $z$ is the comoving coordinate along the $\hat{z}$-axis and
$k$ is the comoving wave-number. In an expanding Universe, the
corresponding {\em physical quantities} are defined as $z_{ph} = z
S(t)$ and $k_{ph} = k/S(t)$, so that $k_{ph} z_{ph} = k z$.

\section{The dispersion relation}

Before discussing the temporal evolution of the perturbation
quantities, it is important to trace what kind of waveforms are
admitted by this system, in the first place. In order to do so, we
have to derive their dispersion relation, $D(k, \om) = 0$, at $t =
t_0$. Provided that certain kinds of modes (such as acoustic,
magnetosonic etc) do exist in the first place, they can be excited
through their interaction with the anisotropic spacetime. An
additional excitation, due to the non-zero resistivity, is also
possible (Fennelly 1980).

Accordingly, we assume a wave-like expansion for the perturbation
quantities, of the form \bea \dl \rho &=& A_{\rh} e^{i(kz - \int^t
\om dt)}, \; \; \; \;
\dl u^{z} = A_u e^{i(kz - \int^t \om dt)} \\
\dl E^{y} &=& A_E e^{i(kz - \int^t \om dt)},\; \; \;
\dl B^{x} = A_B e^{i(kz - \int^t \om dt)} \\
\dl p &=& A_p e^{i(kz - \int^t \om dt)} = C_{s}^{2} A_{\rh}
e^{i(kz - \int^t \om dt)} \eea Although the background quantities
depend on time, in the search for a dispersion relation at $t =
t_0$, we treat the perturbation amplitudes ($A_{i}$s) as {\em
constants}. In this way, our search for potential waveforms at $t
= t_0$, is not disturbed by the inherent non-linearity introduced
for $t > t_0$. Nevertheless, once the potential waveforms are
determined, their interaction with the curved spacetime in the
presence of an external magnetic field, implies that, for $t >
t_0$, the time-dependence of their amplitudes is {\em a priori}
expected. Using Eqs (36), Eq (29) is written in the form \be (H_R
+ 2 H_S - i \om) \dl \rh = - i k \rh(t) \dl u^z \ee where, we have
set \be H_R = {\dot{R} \over R} ~~~ and ~~~ H_S = {\dot{S} \over
S} \ee Furthermore, using Eqs (37), Eq (30) reduce to \be [i \om -
(H_R + H_S + {4 \pi \over \et})] \dl E^y = - i {k \over S} \dl B^x
+ {4 \pi \over \et} S B(t) \dl u^z \ee while, Eq (31) becomes \be
(2 H_S - i \om) \dl B^x = i {k \over S} \dl E^y \ee Finally, Eq
(32) yields \bea & [ & \dot{\rh} (t) + (H_R + 2 H_S -i \om) \rh
(t) ] \dl u^z + i {k \over S^2} C_s^2 \dl \rh = \nn \\ = {1 \over
4 \pi} & [ & {d \over dt} ({B(t) \over S}) + (H_R + 2 H_S - i \om)
{B(t) \over S} ] \dl E^y - \nn \\ & - & i {1 \over 4 \pi} k B(t)
\dl B^x \eea With the aid of Eqs (12) and (14), the combination of
Eqs (39) - (43) results in \bea & [ & - \om^2 + k_{ph}^2 C_s^2 - i
\om (H_R + 2 H_S)] \times \nn \\ & [ & (i \om + H_R + H_S - {4 \pi
\over \et}) (2 H_S - i \om) - k_{ph}^2 ] \nn \\ & = & {4 \pi \over
\et} u_A^2 (H_R + 2 H_S - i \om) \times \nn \\ & [ & (H_R - H_S -i
\om) (2 H _S - i \om) + k_{ph}^2] \eea where, $u_A^2 = (B_0^2 /4
\pi \rh_0)$ is the (dimensionless) Alfv\'en velocity. Eq (44) is
the dispersion relation which determines the possible waveforms
admitted by this dynamical system for every $t \geq t_0$.

We have to point out that, $\om$, as defined by Eqs (34) and (35),
has the usual meaning of the angular frequency of an oscillating
process only in the short-wavelength (high-frequency) regime of
the mode $k$ (Mukhanov, Feldman \& Brandenberger 1992). In other
words, the wave description in curved spacetime makes sense only
when the physical wavelength along the direction of propagation
$[\lm_{ph} = \lm S(t)]$ is much smaller than the corresponding
horizon length $[\ell_{H_S} = H_S^{-1} (t)]$, i.e. \be \lm_{ph}
\ll \ell_{H_S} \ee Eq (45) implies that, in the anisotropic
background (1), the wave description makes sense as long as \be
\om \, , k_{ph} \gg H_R \, , H_S \ee for every $t \geq t_0$. In
this limit, Eq (44) becomes surprisingly transparent, namely \be
(\om^2 - k_{ph}^2 C_s^2) (\om^2 - k_{ph}^2) + i \om {4 \pi \over
\et} \left [\om^2 (1 + u_A^2) - k_{ph}^2 (C_s^2 + u_A^2) \right ]
= 0 \ee Vanishing of the real part results in acoustic $(\om =
k_{ph} C_s)$ and e/m $(\om = k_{ph})$ waves, while, vanishing of
the imaginary part results in fast-magnetosonic waves \be \om^2 (1
+ u_A^2) = k_{ph}^2 (C_s^2 + u_A^2) \ee In fact, in the
zero-resistivity limit (ideal plasma), the obvious modes expected
are the magnetosonic modes, which we recover. On the other hand,
in most astrophysical situations we have (Jackson 1975) \be u_A^2
\ll C_s^2 \ee In this case, Eq (47) reads \be (\om^2 - k_{ph}^2
C_s^2) (\om^2 +i {4 \pi \over \et} \om - k_{ph}^2 ) = 0 \ee
According to Eq (50), in the very high frequency limit where no
acoustic waves are admitted, we are left with a waveform governed
by the dispersion relation \be \om^2 +i {4 \pi \over \et} \om -
k_{ph}^2 = 0 \ee which yields \be e^{-i \int^t \om dt} \sim e^{-
{2 \pi \over \et}t} \ee This result has a clear physical
interpretation: Every very-high-frequency perturbation of the
dynamical system under study, is suppressed due to the finite
resistivity. Therefore, the only modes that survive in a resistive
cosmological model, are the (low-frequency) MHD modes. In the next
Section, we intend to discuss the evolution of these modes.

\section{Numerical study of the MHD mode}

In order to study the temporal evolution of the magnetosonic modes
for $t \geq t_0$, we assume that their amplitudes are no longer
time-independent \bea \dl \rho &=&{\tilde \rho}(t)e^{i(kz - \int^t
\om dt)},\; \; \;
\dl u^{z} ={\tilde u}(t)e^{i(kz - \int^t \om dt)} \\
\dl E^{y}&=&{\tilde E}(t)e^{i(kz - \int^t \om dt)},\;
\dl B^{x}={\tilde B}(t)e^{i(kz - \int^t \om dt)} \\
\dl p&=&{\tilde p}(t)e^{i(kz - \int^t \om dt)}=C_{s}^{2} {\tilde
\rho} (t) e^{i(kz - \int^t \om dt)} \eea In Eqs (53) - (55), the
wave-number $k$ is related to the frequency $\om$ through Eq (48)
and, once again, we have taken into account the equation of state
for the perfect fluid.

We decompose the time-dependent amplitude of the perturbations
(53) - (55) into a real and an imaginary part, as \bea {\tilde
\rho}(t)&=&\rho_{R}(t)
+i \rho_{I}(t)\nn \\
{\tilde u}(t)&=&u_{R}(t)+iu_{I}(t) \\
{\tilde E}(t)&=&E_{R}(t)+iE_{I}(t)\nn \\
{\tilde B}(t)&=&B_{R}(t)+iB_{I}(t) \eea something that reduces Eqs
(29) - (32) to the following first order system \bea
&&\dot{\rh}_{R}+\om \rh_{I} -
\rho_0 ({t_0 \over t})^2 ku_{I}+ {2 \over t} \rho_{R} = 0 \\
&&\dot{\rho }_{I}-\om \rho_{R} + \rho_{0} ({t_0 \over t})^2
ku_{R} + {2 \over t} \rho _{I} = 0 \\
&&\dot{E}_{R}+\om E_{I}+k ({t_0 \over t})^{1/2} B_{I} + {3 \over 2
t} E_{R} + 4 \pi {1 \over \et} E_{R} + \nn \\
&&+ 4 \pi {1 \over \et} B_{0} ({t_0 \over t})^{1/2} u_{R} = 0 \\
&&\dot{E}_{I} - \om E_{R}-k ({t_0 \over t})^{1/2} B_{R} + {3 \over
2 t} E_{I} + 4 \pi {1 \over \et} E_{I} + \nn \\
&&+ 4 \pi {1 \over \et} B_{0} ({t_0 \over t})^{1/2} u_{I} = 0 \\
&&\dot{B}_{R}+\om B_{I} + {1 \over t} B_{R}+ k ({t_0 \over
t})^{1/2} E_{I}=0 \\
&&\dot{B}_{I}-\om B_{R} + {1 \over t} B_{I} - k ({t_0 \over
t})^{1/2} E_{R}=0 \eea \bea &&\rho_{0}\dot{u}_{R} + \om \rh_{0}
u_{I} + ({1 \over 4 \pi t_0} + {1 \over \et}) B_0 ({t_0 \over
t})^{1/2} E_{R} + \nn \\
&&+ {1 \over \et} B_{0}^{2} u_{R} - k C_{s}^{2} ({t \over t_0}) \rh_{I}=0 \\
&&\rho_{0} \dot{u}_{I} - \om \rho_{0} u_{R} + ({1 \over 4 \pi t_0}
+ {1 \over \et}) B_0 ({t_0 \over t})^{1/2} E_{I} + \nn \\
&&+ {1 \over \et} B_{0}^{2} u_{I} + k C_{s}^{2} ({t \over t_0})
\rho_{R}=0 \eea We integrate numerically the system (58) - (65),
using a fifth order Runge-Kutta-Fehleberg scheme with variable
integration step. The time is measured in units of $t_0$ and,
therefore, $\ta = {t \over t_0} \geq 1$. In terms of $\ta$, the
physical wave-number reads $k_{ph} = k/\sqrt{\ta}$ and the Hubble
parameter along the $yz$-plane is written in the form $H_S = (2
\ta t_0)^{-1}$. According to Eq (45), for a certain value of
$\ta$, a wave is well inside the horizon as long as \be k \gg {1
\over 2 \sqrt{\ta} t_0} \ee The validity of Eq (66) for long
$\ta$-intervals determines the appropriate values of the comoving
wave-number. Now, the analysis depends on where do we place the
initial time, $t_0$.

According to the Standard Model (Kolb \& Turner 1990), after
nucleosynthesis, the Universe goes on expanding and cooling but
nothing of great interest takes place until $t \sim 10^{13}$ sec.
At that time, the temperature drops to the point where electrons
and nuclei can form stable atoms (recombination). Before that
time, during the so-called {\em radiation epoch}, photons couple
strongly with matter, the main constituent of which is in the form
of plasma. Therefore, {\em the latest time} at which plasma could
play a role of cosmological significance is the {\em recombination
time} ($t_R = 1.2 \times 10^{13}$ sec). In the limiting case where
$t_0 = t_R$, the condition (66) reads $k \gg {1 \over \sqrt{\ta}}
\times 10^{-14}$ $sec^{-1}$ and, therefore, an appropriate choice
for k would be $k = 10^{-12}$ $sec^{-1}$.

In order to decide on the initial values of the unperturbed
quantities, we write Eq (16) in ordinary units, namely \be \rh_0
c^2 + {B_0^2 \over 8 \pi} = {5 c^2 \over 32 \pi G t_0^2} \ee We
adopt a typical behavior for the energy-density, valid at the late
stages of the radiation epoch (see Weinberg 1972, Eq 15.6.42) \be
\rh_0 c^2 = 1.45 \al T^4 \ee where, $T$ is the temperature and
$\al$ is the black-body constant. At the time of {\em
recombination} ($t_0 = 1.2 \times 10^{13}$ sec, $T = 4000$
$^{\circ} K$), we obtain $\rh_0 \equiv \rh_0 c^2 = 2.8 \;
erg/cm^3$, which, through Eq (67), is effectively a choice on
$B_0$; namely $B_0 \simeq 7$ {\it gauss}. Notice that this value
lies barely within limits of the constraint \be \rh_0 c^2 > {B_0^2
\over 8 \pi} \; , \ee a necessary condition to retain the
anisotropy of the metric (Thorne 1967). Extrapolation of this
result, along the lines of Eq (15), to the present epoch ($t_{p}
\simeq 15 \times 10^9 \; y$), suggests that, today, the
corresponding magnetic field should be $B_{p} \simeq 6.6 \times
10^{-10}$ {\it gauss}. This value lies within limits of the upper
bound for the present-day magnetic field strength, arising from
the large-angular scale anisotropy of the microwave radiation
background (MRB) at last scattering (Barrow, Ferreira \& Silk
1997, Barrow, Maartens \& Tsagas 2006) \be B < 4 \times 10^{-9} \;
\; gauss \ee In fact, now, we may proceed even further, to
estimate the amount of distortion which the expansion anisotropy
along the $x$-axis (caused by the unperturbed magnetic field)
induces to the microwave pattern at the present epoch. The
contribution of a large-scale coherent magnetic field to the
microwave quadrupole anisotropy is given by (Madsen 1989) \be {\Dl
T \over T} \simeq (1 + z)\: {B_{p}^2 \over 8 \pi \eps_{p}} \ee
where, $B_{p}$ and $\eps_{p}$ denote the present values of the
magnetic field and the background radiation energy-density, while,
$z$ is the redshift at which the anisotropy begins to grow (in our
case, at the recombination time, where $z \simeq 1100$). The
present value of the microwave background temperature is $T_{p} =
2.8 \; ^{\circ} K$, corresponding to an energy-density of
$\eps_{p} \simeq 4.7 \times 10^{-13} \; erg/cm^3$ for the
radiation field. Accordingly, our analysis suggests that the
present-day quadrupole anisotropy along the $x$-axis should be \be
\left . {\Dl T \over T} \right \vert_x \simeq 4.06 \times 10^{-5}
\ee i.e. almost four times larger than the corresponding COBE
result.

Taking into account that, initially, the unperturbed quantities
are of the order of unity, we normalize all the perturbation
quantities at $t = t_0$, to $0.01$ in {\em cgs} units. On the
other hand, initially, the equation of state for the perfect fluid
admits $C_s^2 = 0.2$, while, as regards the resistivity, we adopt
the Spitzer relation (Krall \& Trivielpiece 1973) \be \et =
10^{-2} \: ({T \over eV})^{3/2} \; \; \; sec \ee In a
radiation-dominated background, we have (Kolb \& Turner 1990) \be
({T \over eV}) = {10^6 \over \sqrt {t \; (sec)}} \ee and
therefore, during recombination, Eq (64) results in $\et = 0.0645
\; sec$. In order to demonstrate how $\et$ may trigger
instabilities, we consider three cases, namely: $\et = 0.0645 \;
sec$, $\et = 0.0745 \; sec$ and $\et = 0.0870 \; sec$.

The output of the numerical integration consists of the electric
and the magnetic field perturbations' amplitude \bea |\dl E^{y}| &
= & \sqrt{E_{R}^{2}+E_{I}^{2}} \\ |\dl B^{x}| & = &
\sqrt{B_{R}^{2} + B_{I}^{2}} \eea and illustrates their temporal
evolution. In Fig. 1, we present the magnetic field perturbation
versus time. We consider two cases:

\begin{itemize}

\item For $\et = 0$ (ideal plasma), the magnetic field
perturbation grows steeply at early times. It appears that, the
interaction of the perturbed quantities with the anisotropic
spacetime results in the amplification of the {\em convective
field} $\dl E_1^y = - S B(t) \dl u^z$, which is the only one to
survive in the ideal-plasma-limit [e.g. see Eq (30)]. Through
Faraday's law, any amplification in the convective field leads to
an analogous growth in $\dl B^x$, at the expense of the
cosmological expansion. Accordingly, after exhausting any {\em
available} amount of energy, the magnetic field perturbation
reaches at a maximum value, before it is suppressed due to the
cosmological redshift.

\item On the other hand, for $\et \neq 0$, the magnetic field
perturbation also increases rapidly at early times after $t_0$
($\Dl t \sim 10^{11} \; sec$), reaching at values up to 3 times
its initial one. However, in this case, the perturbation's
amplitude is saturated, acquiring sufficiently large values for
long enough time intervals $(\Dl t \sim 10^{12} \; sec)$. This is
due to the fact that, besides the convective field $\dl E_1^y$, a
non-zero resistivity favors also convective currents $(\dl E_2^y =
\et \: \dl J^y)$ (For $\et \neq 0$, the lhs of Eq (30)
corresponds, through Ampere's law, to an electric current).
Accordingly, now, the available energy amount to be absorbed by
the perturbed quantities is larger, and therefore, the magnetic
perturbation remains at high levels for longer time intervals.

As a result, after saturation, the magnitude of $\dl B$
constitutes a fraction of $5 \times 10^{-3}$ of the unperturbed
value of the magnetic strength. In this case, the quadrupole
anisotropy induced in the MRB along the $x$-axis, reads \be \left
. {\Dl T \over T} \right \vert_x \simeq (1 + z) \: {B^2 \over 8
\pi \eps} \: (1 + 2 {\dl B \over B}) \ee resulting in \be \left .
{\Dl T \over T} \right \vert_x \simeq 4.10 \times 10^{-5} \ee i.e.
the corresponding value is enlarged by 1\%.

\end{itemize}

\begin{figure}[h!]
\centerline{\mbox {\epsfxsize=9.cm \epsfysize=7.cm
\epsfbox{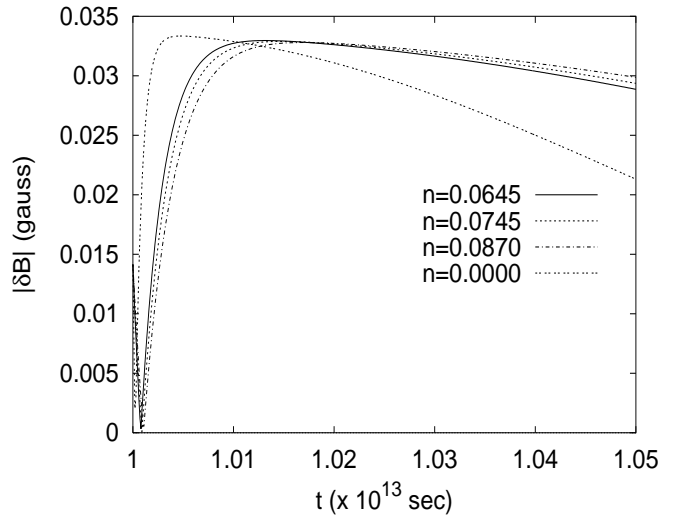}}} \caption{The time-evolution of the
magnetic field perturbation, for $B_0 = 7$ {\it gauss} and for
several values of the resistivity $\et$ $(sec)$. Notice that, for
$\et \neq 0$, the perturbation's amplitude is saturated, acquiring
large values for long enough time intervals.}
\end{figure}

The numerical results indicate a completely different behavior for
the electric field perturbation (Fig. 2). Not only the growth rate
and the highest value of $\dl E^y$ are a little bit smaller than
the corresponding values of $\dl B^x$, but, also, the suppression
rate of the perturbation's amplitude, is much larger than that of
$\dl B^x$, resulting in a rapid decrease of the electric field at
late times. It appears that the expanding Universe disfavors
strong electric fields.

\begin{figure}[h!]
\centerline{\mbox {\epsfxsize=9.cm \epsfysize=7.cm
\epsfbox{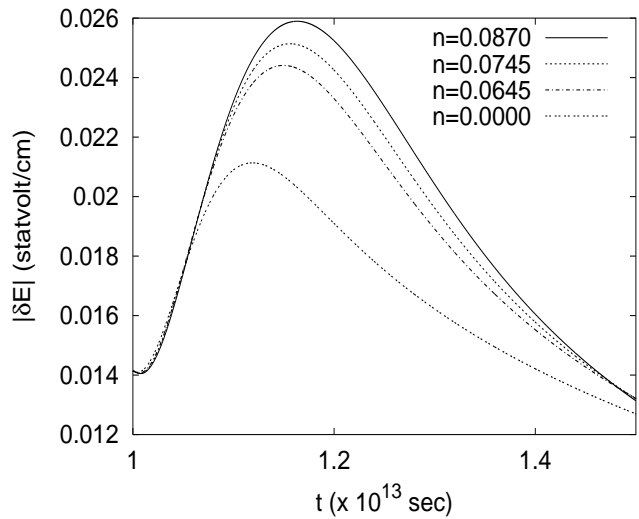}}} \caption{The time-evolution of the
electric field perturbation, for $B_0 = 7$ {\it gauss} and for
several values of the resistivity $\et$ ({\it sec}). Notice that,
in this case, there is no saturation.}
\end{figure}

We conclude that, for reasonable values of the resistivity, the
magnetic field perturbations are lead to a real instability,
acquiring large values for sufficiently long time-intervals. The
influence of resistivity, in triggering instabilities in
anisotropic cosmological models, has been the subject of research
in the past (Fennelly 1980). To the best of our knowledge,
however, this is the first time that a direct connection between
the resistivity and the saturation of the perturbations' amplitude
at high values for long time intervals, is suggested and
discussed.

\section{Jeans-like instabilities}

The question that arises now is, if the cosmological model under
consideration admits also other kinds of instability and which is
their role in connection to the resistive one. To answer this
question, we study the evolution of purely gravitational
perturbations, examining whether they admit a growing behavior
(Jeans instability) or not.

In the absence of e/m fields (and their fluctuations), one is left
with the system of the perturbation equations \bea && (H_R + 2 H_S
-i \om) \dl \rh = -i k \rh(t) \dl u^z \\
&& [ \dot{\rh} (t) + (H_R + 2 H_S - i \om) \rh (t) ] \dl u^z = - i
{k \over S^2} C_s^2 \dl \rh \eea the combination of which, yields
\be [ {\dot{\rh} \over \rh} + (H_R + 2 H_S) - i \om ] [ (H_R + 2
H_S) - i \om ] = - {k^2 \over S^2} C_s^2 \ee Taking into account
the particles' number conservation law, Eq (81) results in \be
\om^2 + i \om (H_R + 2 H_S) - k_{ph}^2 C_s^2 = 0 \ee describing
{\em damped acoustic waves}. With respect to $\om$, Eq (82) is a
second order algebraic equation with roots \be \om_{1,2} = -i {H_R
+2 H_S \over 2} \pm \sqrt{k_{ph}^2 C_s^2 - ({H_R +2 H_S \over
2})^2} \ee provided that \be k_{ph} C_s \geq {H_R +2 H_S \over 2}
\ee In this case, the energy-density perturbations (36) reduce to
\be \dl \rh = {A_{\rh} \over S \sqrt {R}} \: e^{i k z \mp i \int^t
\om_R dt} \ee where, $\om_R$ is given by \be \om_R^2 = {k^2 \over
S^2} C_s^2 - ({H_R + 2 H_S \over 2})^2 \ee Eq (86) represents the
dispersion relation for the propagation of the energy-density
fluctuations. It is worth noting that, in the isotropic case,
where $H_R = H = H_S$, it yields \be \om_R^2 = {k^2 \over S^2}
C_s^2 - {9 \over 4} H^2 \ee which, with the aid of the
corresponding Friedmann equation $H^2 = {8 \pi G \over 3} \rh$,
reads \be \om_R^2 = {k^2 \over S^2} C_s^2 - 6 \pi G \rh \ee Eq
(88) is identical to the isotropic (FRW) result, predicted by
Weinberg (1972), in the relativistic theory of small fluctuations.

In contrast to the high frequency e/m waves (52), as regards the
corresponding energy-density perturbations, propagation is
possible only when their physical wave-number is larger than a
{\em characteristic value}, arising from the condition (84),
otherwise, after some time they become {\em unstable} and grow
exponentially with time (Jeans-like instabilities).

Taking into account the background solution (15), Eq (84) at $t =
t_0$ reads \be k \geq k_c = {1 \over C_s t_0} = 1.86 \times
10^{-13} \; \; cm^{-1} \ee and the corresponding {\em Jeans
length} is given by \be \lm_c = 2 \pi \: C_s \: t_0 = 3.37 \times
10^{13} \; \; cm \ee Propagation of density perturbations with
$\lm > \lm_c$ is not possible, for every $t \geq t_0$ and we are
lead to a {\em gravitational instability}. The larger the
coordinate wave-length is, the most prominent the unstable
behavior will be (Fig. 3).

\begin{figure}[h!]
\centerline{\mbox {\epsfxsize=9.cm \epsfysize=7.cm
\epsfbox{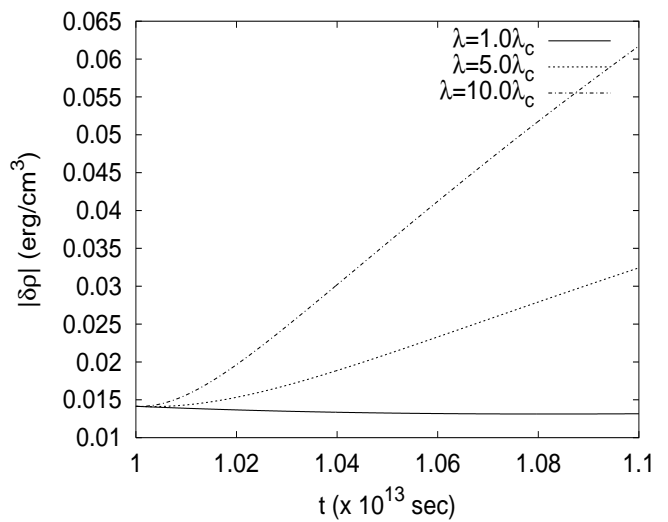}}} \caption{The time-evolution of the
energy density perturbation, for several values of the coordinate
wave-length $\lm$ in terms of $\lm_c$.}
\end{figure}

Furthermore, one may ask whether the waves with wave-number around
$k_c$ in a non-ideal plasma, may grow faster than those in an
ideal plasma. For every $t > t_0$, the {\em physical} Jeans length
along the $\hat{x}$-axis $[\lm_x = \lm_c R(t)]$, is larger than
the corresponding length along the other two axes $[\lm_y = \lm_c
S(t) = \lm_z]$, due to the background anisotropy, suggesting
formation of {\em "cigar-like"} condensations within the
anisotropic fluid. Since this fluid is conductive, these
condensations act in favor of electric currents which may lead to
a further amplification of the e/m perturbations, fortifying any
pre-existing resistive instability (Fig. 4). Therefore, a
Jeans-like instability enhances the phenomena related to the
resistivity.

\begin{figure}[h!]
\centerline{\mbox {\epsfxsize=9.cm \epsfysize=7.cm
\epsfbox{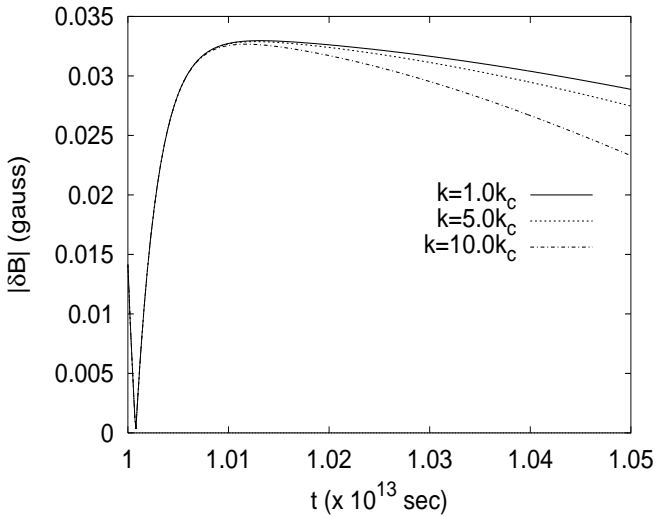}}} \caption{The time-evolution of the
magnetic field perturbation for $\et = 0.0645$ {\it sec}, $B_0 =
7$ {\it gauss} and for several values of the coordinate
wave-number in terms of $k_c$.}
\end{figure}

On the other hand, numerical results indicate that waves with
wavelength around $k_c$ become more prominent as the resistivity
grows (Fig. 5). This result also has a clear physical
interpretation: As we have already seen, any increase in the
resistivity fortifies the surrounding magnetic field. A strong
magnetic field organizes plasma along its lines, favoring any
pre-existing condensations. Hence, resistive instabilities act in
favor of the corresponding gravitational ones and vice versa.

\begin{figure}[h!]
\centerline{\mbox {\epsfxsize=9.cm \epsfysize=7.cm
\epsfbox{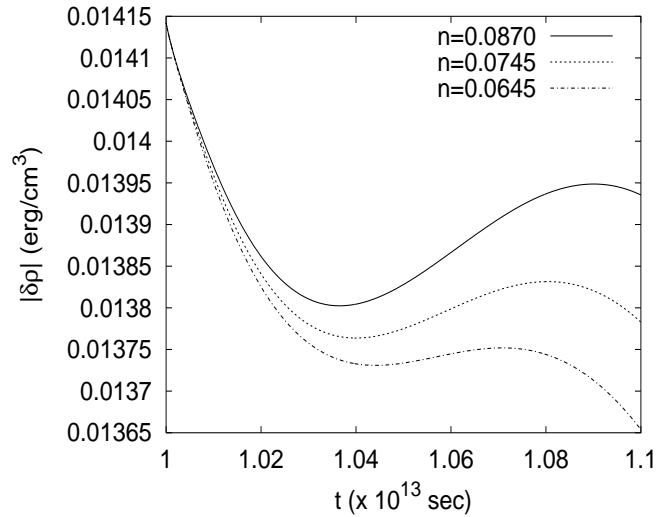}}} \caption{The time-evolution of the
energy density perturbation, for $k = k_c$ and for several values
of the resistivity $\et$ {\it (sec)}. We observe that the Jeans
instability becomes more prominent as the resistivity grows.}
\end{figure}

\section{Discussion}

We study the evolution of the magnetosonic waves in a magnetized,
resistive plasma, which governs the dynamics of an anisotropic
cosmological model. After constructing the general set of MHD and
Einstein's equations for the anisotropic cosmological model (see
the Appendices), we solve the field equations to obtain the
zeroth-order solution. In order to determine the waveforms
admitted by this system in the first place, we introduce wave-like
perturbations and, neglecting all terms higher or equal than the
second order, we extract the dispersion relation at $t = t_0$,
i.e. at the beginning of the interaction between magnetized plasma
and curved spacetime. It appears that magnetosonic modes can be
excited due to the anisotropy and the resistivity. For $t \geq
t_0$, we integrate numerically the perturbed equations, using the
dispersion relation for the fast-magnetosonic waves.

We find that, at early times, both the electric and the magnetic
field perturbations grow exponentially, at least, in the regime
where the linear analysis holds. However there is a major
difference in their behavior in the presence of a non-zero
resistivity. For $\et \neq 0$, the magnetic field perturbation
after increasing to reach at values up to 3 times its initial one,
is subsequently saturated, remaining at high levels for
sufficiently long time intervals $(\Dl t \sim 10^{12} \; sec)$.

The situation is completely different as regards the electric
field perturbation. Not only the growth rate and the highest value
of $\dl E^y$ are a little bit smaller than the corresponding
values of $\dl B^x$, but, also, the suppression rate of the
perturbation's amplitude, is much larger than that of $\dl B^x$.
Accordingly, the electric field decreases rapidly at late times.
It appears that the expanding Universe disfavors strong electric
fields.

Finally, we have shown that waves with wave-number around $k_c$,
are enhanced in non ideal plasmas.

\begin{acknowledgements}
The authors would like to thank Dr Heinz Ishliker and Dr Christos
Tsagas for helpful discussions. Financial support from the Greek
Ministry of Education under the Pythagoras programm, is gratefully
acknowledged.
\end{acknowledgements}

\section*{Appendix A}

We present the closed set of MHD and Einstein equations (in the
system of units where $\hbar = 1 = c$) for the anisotropic
cosmological models of Bianchi-Type I $$ ds^{2} = -dt^2 +
R^{2}(t,z)dx^{2} + S^{2}(t,z)dy^{2} +T^{2}(t,z)dz^{2} \eqno{(A1)}
$$ in the presence of an anisotropic perfect fluid, which allows
for acoustic waves along the ${\hat z}$-direction
$${\cal T}^{\mu \nu }_{fluid} = \rh u^{\mu} u^{\nu} + p_i u^i u^i
+ p_i g^{ii} \eqno{(A2)} $$ and an e/m field of the form $$
F^{\hat{\al} \hat{\bt} }= \left[
\begin{tabular}{cccc}
0& 0& $E^y$&0\\
0& 0& 0& 0\\
$-E^y$& 0& 0& $B^x$\\
0& 0& $-B^x$& 0
\end{tabular}\right] \eqno{(A3)} $$
where, Greek indices refer to the four-dimensional spacetime and
Latin indices refer to the three-dimensional spatial section. In
Eq (A3), $F^{\hat{\al} \hat{\bt} }$ is the Faraday tensor in flat
spacetime. The components of the e/m field in curved spacetime are
defined by $$ F^{\mu \nu} = F^{\hat{\al} \hat{\bt}}
e_{\hat{\al}}^{\mu} e_{\hat{\bt}}^{\nu} \eqno{(A4)}$$ where, the
non-zero components of the orthonormal tetrad
$e_{\hat{\al}}^{\mu}$ of the local Lorentz frame for the metric
(A1), are given by $$ e_{\hat{t}}^{\mu} = (-1, \: 0, \: 0, \: 0)
\;
\; \; e_{\hat{x}}^{\mu} = (0, \: {1 \over R}, \: 0, \: 0) $$ \\
$$ e_{\hat{y}}^{\mu} = (0, \: 0, \: {1 \over S}, \: 0) \;
\; \; e_{\hat{z}}^{\mu} = (0, \: 0, \: 0, \: {1 \over T})
\eqno{(A5)}$$ Therefore, in the curved spacetime (A1), the
non-zero components of the Faraday tensor are $$ F^{02} = {E^y
\over S} = - F^{20} $$ \\ $$ F^{23} = {B^x \over ST} = - F^{32}
\eqno{(A6)}$$ In what follows, the dot denotes time-derivative,
while the prime denotes differentiation with respect to $z$. The
Einstein equations, $G_{\mu \nu} = 8 \pi G ({\cal T}_{\mu
\nu}^{fluid} + {\cal T}_{\mu \nu}^{em})$, result in:

The (tt)-component is given by $$- \frac{1}{T^{2}}
\left(\frac{R^{''}}{R} + \frac{S^{''}}{S}\right) +
\left(\frac{\dot{R}\dot{S}}{R S} + \frac{\dot{S}\dot{T}}{ST} +
\frac{\dot{T}\dot{R}}{T R}\right) - $$ \\
$$ - \frac{1}{T^{2}} \left(\frac{R^{'}S^{'}}{R S} +
\frac{S^{'}T^{'}}{ST}+\frac{T^{'}R^{'}}{T R}\right) = $$ \\
$$ = 8\pi G \rh + G [(E^{y})^{2}+(B^{x})^{2}] \eqno{(A7)} $$

The (xx)-component is given by $$-R^{2} \left( \frac{\ddot{S}}{S}
+\frac{\ddot{T}}{T}\right)+ \frac{R^{2}}{T^{2}}
\left(\frac{S^{''}}{S}\right)-R^{2}
\left(\frac{\dot{S}\dot{T}}{ST}\right)+\frac{R^{2}}{T^{2}}
\left(\frac{T^{'}S^{'}}{T S}\right) = $$ \\
$$= 8 \pi G R^2 p_{x} -
G R^{2}[-(E^{y})^{2}+(B^{x})^{2}] \eqno{(A8)} $$

The (yy)-component is given by $$
-S^{2}\left(\frac{\ddot{R}}{R}+\frac{\ddot{T}}{T}\right)+
\frac{S^{2}}{T^{2}}\left(\frac{R^{''}}{R}\right)-S^{2}
\left(\frac{\dot{R}\dot{T}}{RT}\right)-\frac{S^{2}}{T^{2}}
\left(\frac{T^{'}R^{'}}{T R}\right) = $$ \\
$$= 8 \pi G S^2 p_{y} + G S^2 [-(E^{y})^{2} + (B^{x})^{2}]
\eqno{(A9)} $$

The (zz)-component is given by $$
-T^{2}\left(\frac{\ddot{S}}{S}+\frac{\ddot{R}}{R}\right) +
({T^{\prime \prime} \over T}) - T^{2}
\left(\frac{\dot{S}\dot{R}}{RS}\right) -
\left(\frac{R^{'}S^{'}}{RS}\right)= $$ \\
$$= 8 \pi G T^2 p_{z} + G T^{2}
[+(E^{y})^{2}+(B^{x})^{2}] \eqno{(A10)} $$

The (tz)-component is given by $$
\left(\frac{\dot{R}^{'}}{R}+\frac{\dot{S}^{'}}{S}-\right.
\left.\frac{R^{'}\dot{T}}{RT}-\frac{S^{'}\dot{T}}{ST}\right)=
$$ \\ $$= - 8 \pi G T^{2}\rho u^{z} + 2 G T E^{y} B^{x} \eqno{(A11)} $$

\section*{Appendix B}

On the other hand, the Maxwell equations in curved spacetime are
written in the form
$$F^{\al \bt }_{;\bt }=F^{\al \bt }_{,\bt }+\Gm ^{\bt
}_{\mu \bt } F^{\al \mu }=4\pi J^{\al } $$ \\
$$F_{\al \bt ;\gm }+F_{\bt \gm ;\al }+F_{\gm \al ;\bt }=0 \eqno{(B1)} $$
where, $J^{\al } = {1 \over \et} F^{\al \bt }u_{\bt }$ is the
current density and $\et$ is the electric resistivity of the
fluid. Accordingly, we obtain $$-\partial_{t} E^{y} +
\partial_{z}({ B^{x} \over T}) - E^{y} \left(\frac{\dot{R}}{R} +
\frac{\dot{T}}{T}\right) + $$ \\
$$ + {B^{x} \over T} \left(\frac{R^{'}}{R} + \frac{S^{'}}{S} +
\frac{T^{'}}{T}\right) = 4 \pi {1 \over \et} (u^{0}E^{y}+T
B^{x}u^{z}) \eqno{(B2)} $$ and $$ -\partial_{z} (S E^{y}) +
\partial_{t} (S T B^{x})=0 \eqno{(B3)} $$

\section*{Appendix C}

Taking the {\em time} and {\em space} component of $T_{\; \; ;
\nu}^{\mu \nu} = 0$, we obtain the required equations of motion in
a covariant form, namely $$\partial_{t} [ \rh(t) + {1 \over 8 \pi}
(E^2 + B^2) ] + \partial_z [ \rh u^z - {1 \over 4 \pi T} E B ] + $$\\
$$+ {\dot{R} \over R} \: \left [ p_1(t) - {1 \over 8 \pi} (- E^2 +
B^2) \right ] + {\dot{S} \over S} \: \left [ p_2(t) + {1 \over 8
\pi} (- E^2 + B^2) \right ] + $$ \\
$$ + {\dot{T} \over T} \: \left [ p_3(t) + {1 \over 8 \pi} (+
E^2 + B^2) \right ] + $$ \\
$$+ \left ( {\dot{R} \over R} + {\dot{S} \over S} + {\dot{T} \over
T} \right ) \left [ \rh(t) + {1 \over 8 \pi} (E^2 + B^2) \right ]
= 0 \eqno{(C1)} $$ and $$ \partial_t [ \rh u^z - {1 \over 4 \pi T}
E B ] + \partial_z \lbrace {1 \over T^2} [p_3(t)
+ {1 \over 8 \pi} (E^2 + B^2)] \rbrace - $$ \\
$$ - {R^{\prime} \over R T^2} \left [ p_1(t) - {1 \over 8 \pi} (- E^2 +
B^2) \right ] - $$ \\
$$- {S^{\prime} \over S T^2} \left [ p_2(t) + {1 \over
8 \pi} (- E^2 + B^2) \right ] + $$ \\
$$ + {T^{\prime} \over T^3} \left [ p_3(t) + {1 \over 8 \pi} (
E^2 + B^2) \right ] + $$ \\
$$ + \left ( {\dot{R} \over R} + {\dot{S} \over S} + {\dot{T} \over T}
\right ) \left [ \rh u^z - {1 \over 4 \pi T} E B \right ] + $$
$$+ \left ( {R^{\prime} \over R} + {S^{\prime} \over S} + {T^{\prime}
\over T} \right ) {1 \over T^2} \left [ p_3(t) + {1 \over 8 \pi} (
E^2 + B^2) \right ] = 0 \eqno{(C2)} $$ In addition, the particles'
number conservation law $(\rho u^{\mu })_{;\mu }=0$, results in
$$ \partial_{t}(\rho u^{0})+\partial _{z}(\rho u^{z}) + $$ \\
$$ + \rh u^{0} \left(\frac{\dot{R}}{R} + \frac{\dot{S}}{S} +
\frac{\dot{T}}{T}\right) + \rh u^{z} \left(\frac{R^{'}}{R} +
\frac{S^{'}}{S} + \frac{T^{'}}{T}\right)=0 \eqno{(C3)} $$

\end{document}